\documentclass[11pt]{article}
\usepackage{amssymb}
\usepackage{makeidx}
\usepackage[totalwidth=460truept,totalheight=600truept]{geometry}
\usepackage{latexsym,amssymb,amsmath,graphicx,accents,eucal,slashed,subfigure,stmaryrd}
\usepackage{dsfont}
\usepackage[T1]{fontenc}
\usepackage[hidelinks]{hyperref}

\def\theequation{\arabic{section}.\arabic{equation}}

\renewcommand{\theequation}{\thesection.\arabic{equation}}
\global\arraycolsep=1truept

\numberwithin{equation}{section}
\renewcommand{\theequation}{\arabic{section}.\arabic{equation}}
\newcommand{\ul}[1]{\mkern2mu\underline{\mkern-2mu #1\mkern-2mu}\mkern2mu }

\begin{document}

\bigskip \phantom{C}

\vskip1.4truecm

\begin{center}
{\huge \textbf{Background Field Method}}

\vskip.5truecm

{\huge \textbf{And The Cohomology Of Renormalization}}

\vskip 1truecm

\textsl{Damiano Anselmi}

\vskip .2truecm

\textit{Dipartimento di Fisica ``Enrico Fermi'', Universit\`{a} di Pisa, }

\textit{and INFN, Sezione di Pisa,}

\textit{Largo B. Pontecorvo 3, 56127 Pisa, Italy}

\vskip .2truecm

damiano.anselmi@unipi.it

\vskip 1.5truecm

\textbf{Abstract}
\end{center}

Using the background field method and the Batalin-Vilkovisky formalism, we
prove a key theorem on the cohomology of perturbatively local functionals of
arbitrary ghost numbers, in renormalizable and nonrenormalizable quantum
field theories whose gauge symmetries are general covariance, local Lorentz
symmetry, non-Abelian Yang-Mills symmetries and Abelian gauge symmetries.
Interpolating between the background field approach and the usual,
nonbackground approach by means of a canonical transformation, we take
advantage of the properties of both approaches and prove that a closed
functional is the sum of an exact functional plus a functional that depends
only on the physical fields and possibly the ghosts. The assumptions of the theorem are the
mathematical versions of general properties that characterize the
counterterms and the local contributions to the potential anomalies. This
makes the outcome a theorem on the cohomology of renormalization, rather
than the whole local cohomology. The result supersedes numerous involved
arguments that are available in the literature.

\vfill\eject

\section{Introduction}

\label{s1}

\setcounter{equation}{0}

Locality and gauge invariance allow us to prove that the divergences of
perturbative quantum field theory can be subtracted in a
renormalization-group invariant way, preserving the cancellation of gauge
anomalies to all orders when they vanish at one loop.

Common tricks to handle certain difficulties, e.g. to fix the gauge in a local
way, consist of extending the set of the physical fields $\phi $ to a larger
set $\Phi ^{\alpha }$, which includes the Faddeev-Popov ghosts $C$ \cite%
{faddeev}, the antighosts $\bar{C}$ and suitable Lagrange multipliers $B$
for the gauge fixing. Moreover, to keep track of the effects of
renormalization on the gauge symmetries, external sources $K_{\alpha }$ are
coupled to the transformations of the fields. The extra
fields and the sources simplify several arguments and derivations, but
enlarge the set of counterterms and potential anomalies. It is then
necessary to show that the enlargement has no impact on the physical
quantities.

To ease this task, a canonical formalism is introduced, known as Batalin-Vilkovisky formalism 
\cite{bata}, which collects the Ward-Takahashi-Slavnov-Taylor (WTST)
identities \cite{wtst} in a compact form and generalizes the BRST symmetry 
\cite{brst}. The basic properties of the gauge symmetries are incorporated
into an extended action $S(\Phi ,K)=S_{c}(\phi )-\int R^{\alpha }(\Phi
)K_{\alpha }$, where $S_{c}(\phi )$ is the classical action and the
functions $R^{\alpha }(\Phi )$ are the infinitesimal transformations of the
fields $\Phi ^{\alpha }$. The great advantage of the BV formalism is that
it relates in a simple way the identities satisfied by the action $S(\Phi
,K) $ to the WTST\ identities satisfied by the generating functional $\Gamma 
$ of the one-particle irreducible correlation functions.

A notion of antiparentheses $(X,Y)$ for functionals $X$, $Y$ is defined,
where the fields $\Phi $ and the sources $K$ are viewed as conjugate
variables. Gauge invariance is lifted to a certain identity obeyed by $S$,
the \textit{master equation }$(S,S)=0$, and a certain cohomology. The
counterterms and the local contributions to the potential anomalies are
characterized by being cohomologically closed, i.e. they are local
functionals $X$ that satisfy $(S,X)=0$. A\ local functional is said to be
cohomologically exact, or trivial, if it has the form $(S,Y)$, $Y$ being
another local functional. Two local functionals $X$ and $Y$ are said to be
cohomologically equivalent if $X-Y=(S,Z)$, $Z$ being another local
functional. Throughout this paper, when we speak of local functionals we
include the perturbatively local ones. A perturbatively local functional is
a functional that can be written as a perturbative expansion where each term
of the sum is equal to the spacetime integral of a polynomial function of
the fields, the sources and their derivatives, evaluated in the same point.

To show that the enlargement mentioned above does not extend the set of
cohomological classes, we must prove that the new local functionals $X_{%
\text{new}}$ that can be built with the extra fields and the sources are all
trivial. This problem has been widely studied in the literature.

Kluberg-Stern and Zuber conjectured in ref. \cite{kluberg} that the solution
of $(S,X)=0$ for local functionals $X$ of vanishing ghost number has the
expected form in non-Abelian Yang-Mills theory, i.e. it is the sum of a
local functional $\mathcal{G}(\phi )$ of the physical fields $\phi $ plus a
trivial term $(S,Y)$. Motivated by the Kluberg-Stern--Zuber (KSZ)
conjecture, several people, starting from Dixon and Taylor \cite{dixon} and
Joglekar and Lee \cite{joglekar}, embarked in the brave task of classifying
all the local functionals and operators that are cohomologically closed. A
strong motivation was to work out the most general solutions of the
Wess-Zumino consistency conditions \cite{wesszumino} for the classification
of anomalies \cite{bonora,brandt,sorella}. In refs. \cite{belgi} several
results were generalized and formulated in the context of the BV formalism.
In ref. \cite{coho2} the classification was worked out in detail in the
physically interesting case of Einstein--Yang-Mills theories. For a review
of this approach, see ref. \cite{report}.

Unfortunately, the assumptions under which the Kluberg-Stern--Zuber
conjecture holds are too restrictive. In important cases, including the
standard model, coupled to quantum gravity or not, there exist nontrivial
cohomological classes $X_{\text{new}}(\Phi ,K)$ that depend on the sources $K
$. The reasons are the presence of global symmetries and the $U(1)$ factor
in the gauge group. It is well known that the hypercharges of the matter
fields are not uniquely fixed (up to the overall normalization) by the
tree-level standard-model Lagrangian. The extra terms $X_{\text{new}}$ are
precisely those associated with the free hypercharges. If some counterterms
proportional to $X_{\text{new}}$ were generated by renormalization, they
would jeopardize the cancellation of gauge anomalies. Indeed, the
cancellation of gauge anomalies at one loop imposes further constraints on
the hypercharges and often fixes them uniquely \cite{chargequant}. Thus, it
is crucial to show that renormalization cannot generate the terms $X_{\text{%
new}}$, even if such terms are cohomologically allowed. In several
situations, this result can be achieved with supplementary {\it ad hoc} arguments 
\cite{ABward,ABnonreno}. However, a deeper understanding is most welcome.

It is clear that the cohomology we must consider is not the whole cohomology
of\ the local functionals of the fields and the sources, but the\textit{\
cohomology of renormalization}, that is to say the cohomology of the local
functionals that can be generated by renormalization. This is a sort of
cohomology with constraints. The crucial point is that it is not enough to
characterize the counterterms and the local contributions to the potential
anomalies as being cohomologically closed. Indeed, they satisfy more
restrictive conditions. In this paper, we convert those conditions into
mathematical assumptions and prove a general theorem that bypasses most of
the involved arguments offered so far in the literature and provides a
better understanding of the matter.

In power-counting renormalizable theories, it is relatively easy to list all
the local terms and solve the cohomological problem with a few algebraic
manipulations. On the other hand, in nonrenormalizable theories, such as the
standard model coupled to quantum gravity, the potential counterterms and
the local contributions to anomalies can have arbitrarily large dimensions,
which makes their cohomological classification rather involved. The problem
is equally hard in renormalizable theories, when we include composite fields
of higher dimensions. Most of the arguments that can be found in the
literature are extremely involved and unfit to become part of a quantum
field theory textbook.

The theorem we prove here is simpler and more to the point. We show that the
nontrivial sector of the cohomology of renormalization just depends on the
physical fields $\phi $ and (in the case of functionals of nonvanishing
ghost numbers) the ghosts $C$. Precisely, in general gauge theories whose
gauge symmetries possibly include general covariance, local Lorentz
symmetry, Abelian gauge symmetries and non-Abelian Yang-Mills symmetries,
the solution of the problem $(S,X)=0$, where $X(\Phi ,K)$ is a local
functional generated by renormalization, has the form%
\begin{equation}
\mathcal{G}(\phi ,C)+(S,Y),  \label{theorem}
\end{equation}%
where $\mathcal{G}$ and $Y(\Phi ,K)$ are other local functionals.

The key ingredient of the proof is the use of the background field method 
\cite{backf} and the comparison with the usual, nonbackground approach. The
background field method was formulated in the context of the BV formalism by
Binosi and Quadri in refs. \cite{quadri}\footnote{%
See also ref. \cite{grassi} for a similar approach in the language of WTST\
identities and the Zinn-Justin equation \cite{zinnjustin}.} and by the
present author in ref. \cite{back}. The two approaches differ in some
respects and highlight different properties. Here we take advantage of the
approach of \cite{back}, which offers, in particular, an exhaustive
characterization of the counterterms and the local contributions to the
potential anomalies. Specifically, it allows us to translate the assumption
that $X$ is \textquotedblleft generated by
renormalization\textquotedblright\ into simple mathematical requirements. In
the end, we manage to prove the theorem with a relatively small effort.

The result (\ref{theorem}) implies that the antighosts $\bar{C}$, the
Lagrange multipliers $B$ and the sources $K$ cannot alter the cohomology of
renormalization and offers a better understanding of why renormalization
cannot generate the extra terms $X_{\text{new}}(\Phi ,K)$.

We stress the main differences between the results of this paper and those
of the previous literature, in particular refs. \cite{belgi,coho2,report}.
Those references contain theorems about the \textit{algebraic cohomology} of
local functionals. Precisely, they classify the local solutions $X$ of
the cohomological problem $(S,X)=0$ from the purely algebraic point of
view. In various cases, $K$-dependent solutions $X_{\text{new}}(\Phi ,K)$ are present, and it
is necessary to find {\it ad hoc} arguments to prove that they are
actually not generated by renormalization. Instead, the theorem proved here
goes straight to the point and deals directly with the local functionals
that can be generated by renormalization, which satisfy $(S,X)=0$ plus a few
other assumptions. The $K$-dependent functionals are automatically excluded from the cohomology, and the results are much easier to prove.

We mention an important application of our theorem, to be presented in
detail in a separate publication. The basic tool to prove the cancellation
of gauge anomalies to all orders, once they vanish at one loop, is the
Adler-Bardeen theorem \cite{adlerbardeen,ABrenoYMLR}, which was recently generalized to
nonrenormalizable general gauge theories in ref. \cite{ABnonreno}. A step of
the proof given in \cite{ABnonreno} requires the knowledge of the
cohomological properties satisfied by the counterterms. There, a variant of
the KSZ conjecture for functionals of vanishing ghost number was proved using the results of refs.  \cite{belgi,coho2,report}
and some {\it ad hoc} arguments. Now, it is possible to upgrade the
proof of \cite{ABnonreno} by using the background field method and the
theorem proved here.

The paper is organized as follows. In section \ref{s2} we build the
background field action and work out its relation with the ordinary,
nonbackground action. In section \ref{thetheorem} we state and prove the
theorem, after listing the mathematical assumptions and motivating them
physically. The proof is split into four main steps to make the reader
better appreciate the tricks we use and the guiding philosophy. Section \ref%
{s4} contains the conclusions. The appendix collects some useful formulas
from previous references.

\section{Background field method in the Batalin-Vilkovisky formalism}

\setcounter{equation}{0}

\label{s2}

We assume that the gauge symmetries of the theory are general covariance,
local Lorentz symmetry, Abelian gauge symmetries and non-Abelian Yang-Mills
symmetries, or a subset of these.\ What such symmetries have in common is
that ($i$) the infinitesimal gauge transformations of the physical fields $%
\phi $ are linear functions of $\phi $, and ($ii$) the closure relations are 
$\phi $ independent. These features are crucial, because they endow the
background field method with the renormalization properties we need. An
example of symmetry that does not obey these assumptions is local
supersymmetry.

The Batalin-Vilkovisky formalism is convenient for studying general gauge
theories. It is a type of canonical formalism, where the conjugate variables
are the fields $\Phi ^{\alpha }$ and certain external sources $K_{\alpha }$
coupled to the $\Phi $ transformations. The fields $\Phi ^{\alpha }$ and the
sources $K_{\alpha }$ have statistics $\varepsilon _{\alpha }$ and $%
\varepsilon _{\alpha }+1$, respectively, where $\varepsilon _{\alpha }$ is to 0 mod 2 for
bosons and 1 mod 2 for fermions. A notion of \textit{antiparentheses}%
\begin{equation}
(X,Y)\equiv \int \left( \frac{\delta _{r}X}{\delta \Phi ^{\alpha }}\frac{%
\delta _{l}Y}{\delta K_{\alpha }}-\frac{\delta _{r}X}{\delta K_{\alpha }}%
\frac{\delta _{l}Y}{\delta \Phi ^{\alpha }}\right)  \label{antip}
\end{equation}%
is introduced, where $X$ and $Y$ are functionals of $\Phi $ and $K$, the
subscripts $l$ and $r$ in $\delta _{l}$ and $\delta _{r}$ denote the left
and right functional derivatives, respectively, and the integral is over
spacetime points associated with repeated indices.

The set of fields $\Phi ^{\alpha }=\{\phi ^{i},C^{I},\bar{C}^{I},B^{I}\}$
contains the classical fields $\phi ^{i}$, the Faddeev-Popov ghosts $C^{I}$,
the antighosts $\bar{C}^{I}$ and the Lagrange multipliers $B^{I}$ for the
gauge fixing. The action $S(\Phi ,K)$ is a local functional that solves the
master equation $(S,S)=0$ and coincides with the classical action $%
S_{c}(\phi )$ at $\bar{C}=B=K=0$. The terms that are linear in $K_{\alpha
}=\{K_{\phi }^{i},K_{C}^{I},K_{\bar{C}}^{I},K_{B}^{I}\}$ collect the
infinitesimal transformations $R^{\alpha }(\Phi )$ of the fields. With the gauge
symmetries we are considering here, the master equation admits the simple
solution%
\begin{equation}
\mathcal{S}(\Phi ,K)=S_{c}(\phi )-\int R^{\alpha }(\Phi )K_{\alpha
}=S_{c}(\phi )-\int R_{\phi }^{i}(\phi ,C)K_{\phi }^{i}-\int
R_{C}^{I}(C)K_{C}^{I}-\int B^{I}K_{\bar{C}}^{I},  \label{sfk}
\end{equation}%
which is linear in $K$. Explicit expressions of the functions $R^{\alpha
}(\Phi )$ can be found in the appendix. Note that each $R^{\alpha }(\Phi )$
is at most quadratic in $\Phi $.

Several operations, such as the gauge fixing, can be performed by means of
canonical transformations, which are the transformations $\Phi ,K\rightarrow
\Phi ^{\prime },K^{\prime }$ that preserve the antiparentheses (\ref{antip}%
). They can be derived from a generating functional $F(\Phi ,K^{\prime })$
of fermionic statistics, by means of the formulas%
\begin{equation*}
\Phi ^{\alpha \hspace{0.01in}\prime }=\frac{\delta F}{\delta K_{\alpha
}^{\prime }},\qquad K_{\alpha }=\frac{\delta F}{\delta \Phi ^{\alpha }}.
\end{equation*}%
Given a functional $\chi (\Phi ,K)$ that behaves as a scalar, we write its
transformation law $\chi ^{\prime }(\Phi ^{\prime },K^{\prime })=\chi (\Phi
,K)$ in a compact form as $\chi ^{\prime }=F\chi $.

\subsubsection*{Background field action}

In the framework of the BV formalism, the background field method can be
implemented as follows. Let $\ul{\Phi }$ and $\ul{K}$ denote
the background fields and the background sources. We associate background
fields with just the physical fields $\phi $ and the ghosts $C$, but not the
antighosts $\bar{C}$ and the Lagrange multipliers $B$. Thus, we have $%
\ul{\Phi }^{\alpha }=\{\ul{\phi }^{i},\ul{C}^{I},0,0\}$%
, $\ul{K}_{\alpha }=\{\ul{K}_{\phi }^{i},\ul{K}%
_{C}^{I},0,0\}$ and $R^{\alpha }(\ul{\Phi })=\{R_{\phi }^{i}(%
\ul{\phi },\ul{C}),R_{C}^{I}(\ul{C}),0,0\}$.

One starts from the action%
\begin{equation}
S(\Phi ,K,\ul{\Phi },\ul{K})=S_{c}(\phi )-\int R^{\alpha
}(\Phi )K_{\alpha }-\int R^{\alpha }(\ul{\Phi })\ul{K}_{\alpha
},  \label{sfb}
\end{equation}%
which is obtained from (\ref{sfk}) by adding a background copy with
vanishing classical action. Obviously, the action (\ref{sfb}) satisfies the
master equation%
\begin{equation}
\llbracket  S,S\rrbracket =0,  \label{masts}
\end{equation}%
where the squared antiparentheses are defined as%
\begin{equation*}
\llbracket  X,Y\rrbracket \equiv \int \left( \frac{\delta _{r}X}{\delta \Phi ^{\alpha }}\frac{%
\delta _{l}Y}{\delta K_{\alpha }}-\frac{\delta _{r}X}{\delta K_{\alpha }}%
\frac{\delta _{l}Y}{\delta \Phi ^{\alpha }}+\frac{\delta _{r}X}{\delta 
\ul{\Phi }^{\alpha }}\frac{\delta _{l}Y}{\delta \ul{K}_{\alpha
}}-\frac{\delta _{r}X}{\delta \ul{K}_{\alpha }}\frac{\delta _{l}Y}{%
\delta \ul{\Phi }^{\alpha }}\right)
\end{equation*}%
and act on functionals $X$ and $Y$ of $\Phi $, $K$, $\ul{\Phi }$ and $%
\ul{K}$.

The background shift is the canonical transformation generated by\footnote{%
In this paper, the fields and sources with primes are the transformed ones.
This choice is opposite to the one of \cite{back}, which is why there are
some sign differences with respect to the formulas of that paper.}%
\begin{equation}
F_{\text{b}}(\Phi ,\ul{\Phi },K^{\prime },\ul{K}^{\prime
})=\int (\Phi ^{\alpha }-\ul{\Phi }^{\alpha })K_{\alpha }^{\prime
}+\int \ul{\Phi }^{\alpha }\ul{K}_{\alpha }^{\prime },
\label{fb}
\end{equation}%
which gives the action $F_{\text{b}}S$. The new fields $\Phi ^{\alpha }$ are
called quantum fields and the sources $K_{\alpha }$ are called quantum
sources.

The symmetry transformations $R^{i}(\phi ,C)$ of $\phi ^{i}$ are turned into
the transformations $R^{i}(\phi +\ul{\phi },C+\ul{C})$ of $%
\phi ^{i}+\ul{\phi }^{i}$, which decompose into the sum of

($i$) the background transformations $R^{i}(\ul{\phi },\ul{C})$
of $\ul{\phi }^{i}$, plus

($ii$) the transformations $R^{i}(\phi +\ul{\phi },C+\ul{C}%
)-R^{i}(\ul{\phi },\ul{C})$ of $\phi ^{i}$.

Recalling that the functions $R^{i}(\phi ,C)$ are proportional to $C$, the
transformations ($ii$) split into the sum of

($a$) the quantum transformations $R^{i}(\phi +\ul{\phi },C)$ of $%
\phi ^{i}$, which are given by the $\ul{C}$-independent
contributions, plus

($b$) the background transformations $R^{i}(\phi +\ul{\phi },%
\ul{C})-R^{i}(\ul{\phi },\ul{C})$ of $\phi ^{i}$, which
are given by the $\ul{C}$-dependent contributions.

Something similar happens to the symmetry transformations $R^{I}(C)$ of the
ghosts $C$. Using the fact that the functions $R^{I}(C)$ depend
quadratically on $C$, the quantum transformations of $C^{I}$ are just $%
R^{I}(C)$, and the background transformations of $C^{I}$ are $\int \ul{C}%
^{J}(\delta _{l}R^{I}(C)/\delta C^{J})$, while the background
transformations of $\ul{C}^{I}$ are $R^{I}(\ul{C})$. The
background fields have trivial quantum transformations, because they are
external fields from the quantum field theoretical point of view.

The background transformations of $\bar{C}$ and $B$ remain trivial after the
shift due to $F_{\text{b}}$, which is not what we want. We can adjust them
by making the further transformation generated by%
\begin{equation}
F_{\text{nm}}(\Phi ,\ul{\Phi },K^{\prime },\ul{K}^{\prime
})=\int \Phi ^{\alpha }K_{\alpha }^{\prime }+\int \ul{\Phi }^{\alpha }%
\ul{K}_{\alpha }^{\prime }-\int \mathcal{R}_{\bar{C}}^{I}(\bar{C},%
\ul{C})K_{B}^{I\hspace{0.01in}\prime },  \label{fnm}
\end{equation}%
where $\mathcal{R}_{\bar{C}}^{I}(\bar{C},\ul{C})$ is the antighost
background transformation. More explicitly, the last term of (\ref{fnm})
(together with the minus sign in front of it) is 
\begin{equation}
\int (gf^{abc}\ul{C}^{b}\bar{C}^{c}+\ul{C}^{\rho }\partial
_{\rho }\bar{C}^{a})K_{B}^{a\hspace{0.01in}\prime }+\int (2\ul{C}^{%
\hat{a}\hat{c}}\eta _{\hat{c}\hat{d}}\bar{C}^{\hat{d}\hat{b}}+\ul{C}%
^{\rho }\partial _{\rho }\bar{C}^{\hat{a}\hat{b}})K_{\hat{a}\hat{b}B}^{%
\hspace{0.01in}\prime }+\int \left( \ul{C}^{\rho }\partial _{\rho }%
\bar{C}_{\mu }-\bar{C}_{\rho }\partial _{\mu }\ul{C}^{\rho }\right)
K_{B}^{\mu \hspace{0.01in}\prime },  \label{argu}
\end{equation}%
for Yang-Mills gauge symmetries (including the Abelian ones), local Lorentz
symmetry and diffeomorphisms, where the indices $\hat{a},\hat{b},\ldots $
are local Lorentz indices.

At this point, all the quantum fields transform as matter fields under the background
transformations. For example, a Yang-Mills gauge potential $A_{\mu }^{a}$
behaves as a vector in the adjoint representation, instead of a connection.
Complete formulas of the background transformations are given below.

The theory must be gauge fixed in a background invariant way. This can be
achieved by means of a canonical transformation generated by%
\begin{equation*}
F_{\text{gf}}(\Phi ,\ul{\Phi },K^{\prime },\ul{K}^{\prime
})=\int \ \Phi ^{\alpha }K_{\alpha }^{\prime }+\int \ \ul{\Phi }%
^{\alpha }\ul{K}_{\alpha }^{\prime }-\Psi (\Phi ,\ul{\phi }),
\end{equation*}%
where the gauge fermion $\Psi (\Phi ,\ul{\phi })$ is invariant under
background transformations. Typically, we choose%
\begin{equation}
\Psi (\Phi ,\ul{\phi })=\int \bar{C}^{I}\left( G^{Ii}(\ul{\phi 
},\partial )\phi ^{i}+\zeta _{IJ}(\ul{\phi },\partial )B^{J}\right) ,
\label{psib}
\end{equation}%
where $G^{Ii}(\ul{\phi },\partial )\phi ^{i}$ are the gauge-fixing
functions. The operator matrix $\zeta _{IJ}(\ul{\phi },\partial )$ is
nonsingular at $\ul{\phi }=0$ and symmetric. Invariance
under background transformations can be easily ensured, since $\bar{C}^{I}$
and $B^{I}$ transform as matter fields, while $\ul{\phi }$ and the
plain derivative $\partial $ can be combined into the background covariant
derivative.

For example, we can take%
\begin{eqnarray*}
\Psi &=&\int \sqrt{|\ul{g}|}\left[ \bar{C}^{a}\left( \ul{g}%
^{\mu \nu }\ul{D}_{\mu }A_{\nu }^{a}+\zeta _{ab}B^{b}\right) +\bar{C}%
_{\hat{a}\hat{b}}\left( \ul{e}^{\rho \hat{a}}\ul{g}^{\mu \nu }%
\ul{D}_{\mu }\ul{D}_{\nu }f_{\rho }^{\hat{b}}+\frac{\zeta _{2}%
}{2}B^{\hat{a}\hat{b}}+\frac{\zeta _{3}}{2}\ul{g}^{\mu \nu }%
\ul{D}_{\mu }\ul{D}_{\nu }B^{\hat{a}\hat{b}}\right) \right] \\
&&+\int \sqrt{|\ul{g}|}\bar{C}_{\mu }\left[ \ul{g}^{\mu \nu }%
\ul{g}^{\rho \sigma }\left( \ul{D}_{\rho }h_{\sigma \nu
}+\zeta _{4}\ul{D}_{\nu }h_{\rho \sigma }\right) +\frac{\zeta _{5}}{2}%
\ul{g}^{\mu \nu }B_{\nu }\right] .
\end{eqnarray*}%
Here $\ul{A}_{\mu }^{a}$, $\ul{e}_{\mu }^{a}$ and $\ul{g%
}_{\mu \nu }$ denote the background gauge fields, vielbein and metric,
respectively, while $A_{\mu }^{a}$, $f_{\mu }^{a}$ and $h_{\mu \nu }$ are
the respective quantum fluctuations. The tensor $\zeta _{ab}$ is constant
and proportional to the identity in every simple subgroup of the Yang-Mills
gauge group, while $\zeta_i$ are other constants. Finally, $\ul{D}$
denotes the covariant derivative on the background fields.

The three canonical transformations listed so far can be easily composed
using the theorems of ref. \cite{CBHcanon}, recalled in the appendix. In our
case, it suffices to use formula (\ref{thispa}), because the nontrivial
sectors of the canonical transformations do not contain any pairs of
conjugate variables besides $B$ and $K_{B}^{\prime }$, on which they depend
linearly. The corrections to (\ref{thispa}) contain second or higher
derivatives, as shown in (\ref{fista}), so they vanish. The composition
gives the generating functional%
\begin{equation*}
F_{\text{gf}}F_{\text{nm}}F_{\text{b}}=\int (\Phi ^{\alpha }-\ul{\Phi 
}^{\alpha })K_{\alpha }^{\prime }+\int \ul{\Phi }^{\alpha }\ul{%
K}_{\alpha }^{\prime }-\int \mathcal{R}_{\bar{C}}^{I}(\bar{C},\ul{C}%
)K_{B}^{I\hspace{0.01in}\prime }-\Psi (\Phi -\ul{\Phi },\ul{%
\phi })+\int \bar{C}^{I}\zeta _{IJ}(\ul{\phi },\partial )\mathcal{R}_{%
\bar{C}}^{J}(\bar{C},\ul{C}).
\end{equation*}

Applying this canonical transformation to the action (\ref{sfb}), we obtain
the background field gauge-fixed action 
\begin{equation}
S_{\text{b}}=F_{\text{gf}}F_{\text{nm}}F_{\text{b}}S,  \label{esb}
\end{equation}%
which can be decomposed as the sum of a quantum action $\hat{S}_{\text{b}}$
plus a background action $\bar{S}_{\text{b}}$. Precisely, we have 
\begin{equation}
S_{\text{b}}(\Phi ,\ul{\Phi },K,\ul{K})=\hat{S}_{\text{b}%
}(\Phi ,\ul{\phi },K)+\bar{S}_{\text{b}}(\Phi ,\ul{\Phi },K,%
\ul{K})\text{,}  \label{sb}
\end{equation}%
where%
\begin{eqnarray}
\hat{S}_{\text{b}}(\Phi ,\ul{\phi },K) &=&S_{c}(\phi +\ul{\phi 
})-\int R^{\alpha }(\phi +\ul{\phi },C,\bar{C},B)\tilde{K}_{\alpha },
\label{shat} \\
\bar{S}_{\text{b}}(\Phi ,\ul{\Phi },K,\ul{K}) &=&-\int 
\mathcal{R}^{\alpha }(\Phi ,\ul{C})K_{\alpha }-\int R^{\alpha }(%
\ul{\Phi })\ul{K}_{\alpha }.  \label{sbarra}
\end{eqnarray}%
Here the tilde sources $\tilde{K}_{\alpha }$ coincide with $K_{\alpha }$
apart from $\tilde{K}_{\phi }^{i}$ and $\tilde{K}_{\bar{C}}^{I}$, which are
given by 
\begin{equation*}
\tilde{K}_{\phi }^{i}=K_{\phi }^{i}-\bar{C}^{I}G^{Ii}(\ul{\phi },-%
\overleftarrow{\partial }),\qquad \tilde{K}_{\bar{C}}^{I}=K_{\bar{C}%
}^{I}-G^{Ii}(\ul{\phi },\partial )\phi ^{i}-\zeta _{IJ}(\ul{%
\phi },\partial )B^{J}.
\end{equation*}

The source-dependent sectors of the actions $\hat{S}_{\text{b}}$ and $\bar{S}%
_{\text{b}}$ encode the quantum transformations and the background
transformations, respectively. The curly $R$ denotes the background
transformations of the quantum fields. Those of the antighosts are $%
\mathcal{R}_{\bar{C}}^{I}$, and those of $\phi $ and $C$ are given by the
formula 
\begin{equation}
\mathcal{R}^{\alpha }(\Phi ,\ul{C})=R^{\alpha }(\Phi +\ul{\Phi 
})-R^{\alpha }(\ul{\Phi })-R^{\alpha }(\phi +\ul{\phi },C,\bar{%
C},B),  \label{boc}
\end{equation}%
while those of the Lagrange multipliers $B$ are given by%
\begin{equation}
\mathcal{R}_{B}^{I}(B,C)=-\int B^{J}\frac{\delta _{l}}{\delta \bar{C}^{J}}%
\mathcal{R}_{\bar{C}}^{I}(\bar{C},C).  \label{iddi}
\end{equation}%
We recall that the transformations $R^{\alpha }$ and $\mathcal{R}^{\alpha }$
obey the identity \cite{back}%
\begin{equation}
\int \left( R_{C}^{J}\frac{\delta _{l}}{\delta C^{J}}+\mathcal{R}_{\bar{C}%
}^{J}(\bar{C},C)\frac{\delta _{l}}{\delta \bar{C}^{J}}\right) \mathcal{R}_{%
\bar{C}}^{I}(\bar{C},C)=0,  \label{iddo}
\end{equation}%
which can be easily checked in the three cases (\ref{argu}) and is useful to
work out the formulas (\ref{shat}) and (\ref{sbarra}).

For example, in the case of Yang-Mills symmetry the gauge fields $A_{\mu
}^{a}$ have $R_{\mu }^{a}(\Phi )=\partial _{\mu }C^{a}+gf^{abc}A_{\mu
}^{b}C^{c}$, so formula (\ref{boc}) gives $\mathcal{R}_{\mu }^{a}(\Phi ,%
\ul{C})=-gf^{abc}\ul{C}^{b}A_{\mu }^{c}$. The ghosts $C^{a}$
have $R_{C}^{a}(\Phi )=-(g/2)f^{abc}C^{b}C^{c}$, so $\mathcal{R}%
_{C}^{a}(\Phi ,\ul{C})=-gf^{abc}\ul{C}^{b}C^{c}$.

Clearly, (\ref{masts}) and (\ref{esb}) imply that the background field
action satisfies the master equation%
\begin{equation}
\llbracket  S_{\text{b}},S_{\text{b}}\rrbracket =0,  \label{mastb}
\end{equation}%
which splits into%
\begin{equation}
\llbracket  \hat{S}_{\text{b}},\hat{S}_{\text{b}}\rrbracket =\llbracket  \hat{S}_{\text{b}},\bar{S}_{%
\text{b}}\rrbracket =\llbracket  \bar{S}_{\text{b}},\bar{S}_{\text{b}}\rrbracket =0.  \label{mastbsplit}
\end{equation}%
To prove the splitting, rescale $\ul{C}$ and $\ul{K}_{C}$ by $%
\tau $ and $1/\tau $, respectively. Then, $\hat{S}_{\text{b}}\rightarrow 
\hat{S}_{\text{b}}$ and $\bar{S}_{\text{b}}\rightarrow \tau \bar{S}_{\text{b}%
}$, so $\llbracket  S_{\text{b}},S_{\text{b}}\rrbracket $ is a quadratic polynomial in $\tau $%
. Setting the three coefficients of the polynomial to zero gives (\ref%
{mastbsplit}).

A way to characterize a functional $X$ that is invariant under the
background transformations is to say that it satisfies $\llbracket  \bar{S}_{\text{b}%
},X\rrbracket =0$. In particular, the gauge fermion (\ref{psib}) must satisfy $\llbracket  
\bar{S}_{\text{b}},\Psi \rrbracket =0$.

\subsubsection*{Nonbackground action}

The nonbackground gauge-fixed action is $F_{\text{gf}}^{\prime }S$, where%
\begin{equation}
F_{\text{gf}}^{\prime }(\Phi ,\ul{\Phi },K^{\prime },\ul{K}%
^{\prime })=\int \ \Phi ^{\alpha }K_{\alpha }^{\prime }+\int \ \ul{%
\Phi }^{\alpha }\ul{K}_{\alpha }^{\prime }-\Psi ^{\prime }(\Phi ),
\label{fgfp}
\end{equation}%
is the canonical transformation that performs the gauge fixing. The
background fields and sources are inert here. We have included them just for
comparison with the background field action. To simplify the
renormalization, we take a quadratic gauge fermion $\Psi ^{\prime }$, such as%
\begin{equation*}
\Psi ^{\prime }(\Phi )=\int \bar{C}^{I}\left( G^{Ii}(0,\partial )\phi
^{i}+\zeta _{IJ}(0,\partial )B^{J}\right) .
\end{equation*}

For convenience, we define the nonbackground action $S_{\text{nb}}$ by
making a further background shift through the transformation $F_{\text{b}}$.
So doing, we have 
\begin{equation}
S_{\text{nb}}=F_{\text{b}}F_{\text{gf}}^{\prime }S.  \label{esnb}
\end{equation}%
We easily find the decomposition%
\begin{equation}
S_{\text{nb}}(\Phi ,\ul{\Phi },K,\ul{K})=\hat{S}_{\text{nb}%
}(\Phi +\ul{\Phi },K)+\bar{S}_{\text{nb}}(\Phi ,\ul{\Phi },K,%
\ul{K}),  \label{snb}
\end{equation}%
where%
\begin{eqnarray}
\hat{S}_{\text{nb}}(\Phi +\ul{\Phi },K) &=&S_{c}(\phi +\ul{%
\phi })-\int R^{\alpha }(\Phi +\ul{\Phi })\bar{K}_{\alpha },
\label{shatp} \\
\bar{S}_{\text{nb}}(\ul{\Phi },K,\ul{K}) &=&-\int R^{\alpha }(%
\ul{\Phi })(\ul{K}_{\alpha }-K_{\alpha }),  \label{sbarp}
\end{eqnarray}%
and the sources $\bar{K}_{\alpha }$ with bars coincide with $K_{\alpha }$
apart from $\bar{K}_{\phi }^{i}$ and $\bar{K}_{\bar{C}}^{I}$, which are 
\begin{equation}
\bar{K}_{\phi }^{i}=K_{\phi }^{i}-\bar{C}^{I}G^{Ii}(0,-\overleftarrow{%
\partial }),\qquad \bar{K}_{\bar{C}}^{I}=K_{\bar{C}}^{I}-G^{Ii}(0,\partial
)(\phi ^{i}+\ul{\phi }^{i})-\zeta _{IJ}(0,\partial )B^{J}.
\label{kbar}
\end{equation}

The nonbackground action satisfies the master equation%
\begin{equation}
\llbracket  S_{\text{nb}},S_{\text{nb}}\rrbracket =0,  \label{mastnb}
\end{equation}%
which splits into%
\begin{equation}
\llbracket  \hat{S}_{\text{nb}},\hat{S}_{\text{nb}}\rrbracket =\llbracket  \hat{S}_{\text{nb}},\bar{S}_{%
\text{nb}}\rrbracket =\llbracket  \bar{S}_{\text{nb}},\bar{S}_{\text{nb}}\rrbracket =0.
\label{mastnbsplit}
\end{equation}%
The splitting is a direct consequence of (\ref{shatp}) and (\ref{sbarp}).

\subsubsection*{Interpolation between the background and nonbackground
actions}

To switch back and forth between the background and nonbackground approaches
we must make the canonical transformation $F_{\text{b}}F_{\text{gf}}^{\prime
}F_{\text{b}}^{-1}F_{\text{nm}}^{-1}F_{\text{gf}}^{-1}$, since formulas (\ref%
{esb}) and (\ref{esnb}) give%
\begin{equation*}
S_{\text{nb}}=F_{\text{b}}F_{\text{gf}}^{\prime }F_{\text{b}}^{-1}F_{\text{%
nm}}^{-1}F_{\text{gf}}^{-1}S_{\text{b}}.
\end{equation*}%
Using again formula (\ref{thispa}), we easily find the generating function%
\begin{equation}
F_{\text{b}}F_{\text{gf}}^{\prime }F_{\text{b}}^{-1}F_{\text{nm}}^{-1}F_{%
\text{gf}}^{-1}=\int \Phi ^{\alpha }K_{\alpha }^{\prime }+\int \ \ul{%
\Phi }^{\alpha }\ul{K}_{\alpha }^{\prime }+\Delta \Psi (\Phi ,%
\ul{\Phi })+\int \mathcal{R}_{\bar{C}}^{I}(\bar{C},\ul{C}%
)K_{B}^{I\hspace{0.01in}\prime }-\int \bar{C}^{I}\zeta _{IJ}(0,\partial )%
\mathcal{R}_{\bar{C}}^{J}(\bar{C},\ul{C}),  \label{gives}
\end{equation}%
where 
\begin{equation*}
\Delta \Psi (\Phi ,\ul{\Phi })=\int \bar{C}^{I}\left[ G^{Ii}(%
\ul{\phi },\partial )\phi ^{i}-G^{Ii}(0,\partial )(\phi ^{i}+%
\ul{\phi }^{i})+(\zeta _{IJ}(\ul{\phi },\partial )-\zeta
_{IJ}(0,\partial ))B^{J}\right] .
\end{equation*}

It is convenient to express the canonical transformation (\ref{gives}) by
means of the componential map $\mathcal{C}$ of ref. \cite{CBHcanon}, also
recalled in the appendix. We can write%
\begin{equation*}
F_{\text{b}}F_{\text{gf}}^{\prime }F_{\text{b}}^{-1}F_{\text{nm}}^{-1}F_{%
\text{gf}}^{-1}=\mathcal{C}(Q),
\end{equation*}%
where%
\begin{equation}
Q(\Phi ,\ul{\Phi },K^{\prime })=\Delta \Psi (\Phi ,\ul{\Phi }%
)+\int \mathcal{R}_{\bar{C}}^{I}(\bar{C},\ul{C})K_{B}^{I\hspace{0.01in%
}\prime }-\frac{1}{2}\int \bar{C}^{I}\left[ \zeta _{IJ}(\ul{\phi }%
,\partial )+\zeta _{IJ}(0,\partial )\right] \mathcal{R}_{\bar{C}}^{J}(\bar{C}%
,\ul{C}).  \label{Q}
\end{equation}%
Again, this functional does not contain any pairs of conjugate variables
besides $B$ and $K_{B}^{\prime }$, and depends linearly on them. Thus, the
expansion (\ref{expac}) effectively reduces to (\ref{boso}), which gives (%
\ref{gives}).

We can continuously interpolate between the two approaches by introducing a
parameter $\xi $ that varies from $0$ to $1$ and make the canonical
transformation generated by $\mathcal{C}(\xi Q)$, whose inverse is $\mathcal{%
C}(-\xi Q)$. We find 
\begin{eqnarray}
\mathcal{C}(\xi Q) &=&\int \ \Phi ^{\alpha }K_{\alpha }^{\prime }+\int \ 
\ul{\Phi }^{\alpha }\ul{K}_{\alpha }^{\prime }+\xi \Delta \Psi
+\xi \int \mathcal{R}_{\bar{C}}^{I}(\bar{C},\ul{C})K_{B}^{I\hspace{%
0.01in}\prime }  \notag \\
&&-\frac{\xi }{2}\int \bar{C}^{I}\left[ (1-\xi )\zeta _{IJ}(\ul{\phi }%
,\partial )+(1+\xi )\zeta _{IJ}(0,\partial )\right] \mathcal{R}_{\bar{C}%
}^{J}(\bar{C},\ul{C}).  \label{passo}
\end{eqnarray}

The form of this transformation is important to simplify some arguments that
follow. We define the interpolating action as%
\begin{equation}
S_{\xi }=\mathcal{C}(-\xi Q)S_{\text{nb}}=\mathcal{C}(-\xi Q)F_{\text{b}}F_{%
\text{gf}}^{\prime }S.  \label{inter}
\end{equation}%
Explicitly, we have 
\begin{equation}
S_{\xi }(\Phi ,\ul{\Phi },K,\ul{K})=S_{c}(\phi +\ul{%
\phi })-\int R^{\alpha }(\Phi +\ul{\Phi })\tilde{K}_{\alpha }(\xi
)-\xi \int \mathcal{R}_{\bar{C}}^{I}(\bar{C},\ul{C})\tilde{K}_{\bar{C}%
}^{I}(\xi )-\int R^{\alpha }(\ul{\Phi })(\ul{\tilde{K}}%
_{\alpha }(\xi )-\tilde{K}_{\alpha }(\xi )),  \label{sxi}
\end{equation}%
where $\tilde{K}_{C}^{I}(\xi )=K_{C}^{I}$, 
\begin{equation}
\tilde{K}_{\phi }^{i}(\xi )=K_{\phi }^{i}-\xi \bar{C}^{I\hspace{0.01in}%
}G^{Ii}(\ul{\phi },-\overleftarrow{\partial })-(1-\xi )\bar{C}^{I%
\hspace{0.01in}}G^{Ii}(0,-\overleftarrow{\partial }),  \label{ktfi}
\end{equation}%
while the other $\xi $-dependent tilde sources have expressions that we do
not need to report here. It suffices to say that they are linear functions
of the quantum fields $\Phi $, apart from $\ul{\tilde{K}}_{\phi
}^{i}(\xi )$ and $\ul{\tilde{K}}_{C}^{I}(\xi )$, which are quadratic.
Moreover, the differences $\tilde{K}_{\alpha }(\xi )-K_{\alpha }$ and $%
\ul{\tilde{K}}_{\alpha }(\xi )-\ul{K}_{\alpha }$ are
independent of the sources $K$ and $\ul{K}$ other than $K_{B}^{I}$.
In particular, $\delta _{r}S_{\xi }/\delta \ul{K}_{\alpha
}=-R^{\alpha }(\ul{\Phi })$. Note that the $B$-dependent terms of the
interpolating action (\ref{sxi}) are quadratic functions of the quantum
fields. Moreover, the $K_{\bar{C}}^{I}$-dependent terms are just%
\begin{equation}
-\int \left( B^{I}+\xi \mathcal{R}_{\bar{C}}^{I}(\bar{C},\ul{C}%
)\right) K_{\bar{C}}^{I}  \label{chio}
\end{equation}%
and the $K_{B}$-dependent terms are linear in the quantum fields.

Obviously, $S_{\xi }$ satisfies the master equation%
\begin{equation}
\llbracket  S_{\xi },S_{\xi }\rrbracket =0,  \label{sm1}
\end{equation}%
which is a direct consequence of (\ref{masts}) and (\ref{inter}).

\section{The theorem}

\setcounter{equation}{0}

\label{thetheorem}

In this section we state and prove the main theorem of this paper. Since the
parameter $\xi $ is introduced by means of the canonical transformation (\ref%
{passo}), the action $S_{\xi }$ satisfies the differential equation%
\begin{equation}
\frac{\partial S_{\xi }}{\partial \xi }-\llbracket  S_{\xi },\tilde{Q}\rrbracket =0,
\label{i0}
\end{equation}%
where $\tilde{Q}(\Phi ,\ul{\Phi },K)$ coincides with $Q(\Phi ,%
\ul{\Phi },K)$. Note that the sources have no primes in the last
expression. We use different symbols for the two functionals $\tilde{Q}$ and 
$Q$, since the natural variables of $\tilde{Q}$ are the fields and the
sources without primes, while the natural variables of $Q$ are the fields
without primes and the sources with primes.

The result (\ref{i0}) can be proved as follows. The nonbackground action $S_{%
\text{nb}}=\mathcal{C}(\xi Q)S_{\xi }$ obviously satisfies $\partial S_{%
\text{nb}}/\partial \xi =0$. By formula (\ref{bu}), recalled in the
appendix, the transformed action $S_{\xi }$ satisfies (\ref{i0}) as long as $%
\tilde{Q}$ coincides with the derivative $\partial \mathcal{C}(\xi
Q)/\partial \xi $, after the sources with primes are expressed in terms of
the fields and the sources without primes. This fact, which can be checked
directly in our case, is a general property of the componential map \cite%
{CBHcanon}. Indeed, $\tilde{Q}$ plays the role of the Hamiltonian associated
with a fictitious time evolution parametrized by $\xi $ and $\mathcal{C}(\xi
Q)$ solves the Hamilton-Jacobi equation.

\subsubsection*{Assumptions}

We consider local functionals $X_{\xi }$ and $Y_{\xi }$ of $\Phi $, $K$ and $%
\ul{\Phi }$ that satisfy the equations%
\begin{eqnarray}
\llbracket  S_{\xi },X_{\xi }\rrbracket  &=&0,  \label{i1} \\
\frac{\partial X_{\xi }}{\partial \xi }-\llbracket  X_{\xi },\tilde{Q}\rrbracket  &=&\llbracket  S_{\xi
},Y_{\xi }\rrbracket .  \label{i2}
\end{eqnarray}

We use the notation $X_{0}$, $X_{1}$ to denote the functional $X_{\xi }$ at $%
\xi =0$ and $\xi =1$, respectively. We assume that

($i$) $X_{0}$ just depends on $\Phi +\ul{\Phi }$ and $K$.

($ii$) $X_{1}$ just depends on $\ul{\Phi }$ at $\phi =C=0$.

($iii$) $X_{\xi }$ and $Y_{\xi }$ are independent of $B$, $K_{\bar{C}}$ and $%
K_{B}$.

Let $g$ denote the ghost number of $X_{\xi }$. Then, $Y_{\xi }$ has ghost
number $g-1$. For future use, we show that assumption ($ii$) can be replaced
by

($ii^{\prime }$) $X_{1}$ is a sum of terms that contain $g$ background
ghosts $\ul{C}$ at $\phi =C=0$.

When we set $\phi =C=0$, no objects of positive ghost number remain, besides 
$\ul{C}$. If ($ii^{\prime }$) holds, we can drop all objects that
have negative ghost numbers inside $X_{1}$, which means $\bar{C}$ and all
the sources $K$ except $K_{\bar{C}}$. Since ($iii$) tells us that $X_{1}$
does not depend on $B$ and $K_{\bar{C}}$, ($ii$) follows. On the other hand,
if ($ii$) holds, $X_{1}$ just depends on $\ul{\phi }$ and $\ul{%
C}$ at $\phi =C=0$. Then it must be a sum of terms that contain $g$
background ghosts $\ul{C}$, because it has ghost number $g$. This
implies ($ii^{\prime }$).

Now we explain the meaning of the assumptions listed above. In most
applications, the functionals $X_{\xi }$ and $Y_{\xi }$ are originated by
the counterterms and the local contributions to the potential anomalies. In
particular, in the case $g=0$ the functional $X_{\xi }$ typically comes from
the renormalization of the action, while $Y_{\xi }$ comes from the
renormalization of the average $\langle \tilde{Q}\rangle $.

Without entering into details, we recall that the formulas (\ref{i1}) and (%
\ref{i2}) are typical consequences of the regularized and partially
renormalized versions of the equations of gauge invariance and gauge
independence \cite{ABward}, which are%
\begin{equation}
\llbracket  \Gamma _{R},\Gamma _{R}\rrbracket =0,\qquad \frac{\partial \Gamma _{R}}{\partial
\xi }-\llbracket  \Gamma _{R},\langle \tilde{Q}_{R}\rangle \rrbracket =0,  \label{gauind}
\end{equation}%
where $\Gamma _{R}$ is the renormalized $\Gamma $ functional and $\tilde{Q}%
_{R}$ is the renormalized $\tilde{Q}$. When $\Gamma _{R}$ and $\tilde{Q}_{R}$
are just partially renormalized, say up to and including $n$ loops, the
equations of gauge invariance and gauge independence give conditions on the $%
(n+1)$-loop counterterms, which have the forms (\ref{i1}) and (\ref{i2}).

Condition ($i$) is dictated by the properties of the nonbackground action (%
\ref{snb}), encoded in the formulas (\ref{sbarp}) and (\ref{shatp}). If the
regularized action $S_{\text{nb}}^{\text{reg}}$ has a similar structure,
that is to say%
\begin{equation*}
S_{\text{nb}}^{\text{reg}}(\Phi ,\ul{\Phi },K,\ul{K})=\hat{S}_{%
\text{nb}}^{\text{reg}}(\Phi +\ul{\Phi },K)+\bar{S}_{\text{nb}}(%
\ul{\Phi },K,\ul{K}),
\end{equation*}%
where $\bar{S}_{\text{nb}}$ is unmodified, then the renormalized action also
has this structure. Note that $\bar{S}_{\text{nb}}$ is just made of external
fields, so it cannot contribute to the nontrivial Feynman diagrams.
Moreover, the counterterms do not depend on $\Phi $ and $\ul{\Phi }$
separately, but only on their sum, as specified in ($i$). Finally, $\bar{S}_{%
\text{nb}}$ is not renormalized, for this very reason.

Similarly, conditions ($ii$) and ($ii^{\prime }$) are dictated by the
properties of the background field side. Indeed, we know that the action $S_{%
\text{b}}$ of formula (\ref{sb}) splits into (\ref{shat}) plus (\ref{sbarra}%
). If the regularized action $S_{\text{b}}^{\text{reg}}$ has a similar
structure, that is to say%
\begin{equation*}
S_{\text{b}}^{\text{reg}}(\Phi ,\ul{\Phi },K,\ul{K})=\hat{S}_{%
\text{b}}^{\text{reg}}(\Phi ,\ul{\phi },K)+\bar{S}_{\text{b}}(\Phi ,%
\ul{\Phi },K,\ul{K}),
\end{equation*}%
the counterterms just depend on $\Phi $, $\ul{\phi }$ and $K$, so
they satisfy property ($ii^{\prime }$) for $g=0$. Note that $\bar{S}_{\text{b}%
} $, which is linear in the quantum fields, does not contribute to any
nontrivial one-particle irreducible diagrams. It is also not renormalized,
because it vanishes at $\ul{C}=0$, while the counterterms are
independent of $\ul{C}$.

The case $g=1$ is also important, because it concerns the potential gauge
anomalies. When the dimensional regularization -- or any
regularization that embeds the dimensional one -- is used, the gauge anomalies are
encoded in the functional $\langle \llbracket  S_{\text{b}}^{\text{reg}},S_{\text{b}%
}^{\text{reg}}\rrbracket \rangle =\langle \llbracket  \hat{S}_{\text{b}}^{\text{reg}},\hat{S}_{%
\text{b}}^{\text{reg}}\rrbracket \rangle +2\langle \llbracket  \hat{S}_{\text{b}}^{\text{reg}},%
\bar{S}_{\text{b}}\rrbracket \rangle $ (for the proof see, for example, the appendix
of ref. \cite{ABrenoYMLR}), so they are linear functions of $\ul{C}$.
Indeed, $\langle \llbracket  \hat{S}_{\text{b}}^{\text{reg}},\hat{S}_{\text{b}}^{%
\text{reg}}\rrbracket \rangle $ is $\ul{C}$ independent, while $\langle \llbracket  
\hat{S}_{\text{b}}^{\text{reg}},\bar{S}_{\text{b}}\rrbracket \rangle $ contains one
power of $\ul{C}$. Thus, $\langle \llbracket  S_{\text{b}}^{\text{reg}},S_{%
\text{b}}^{\text{reg}}\rrbracket \rangle $ must be proportional to $\ul{C}$ at 
$\phi =C=0$, because it has ghost number 1 and no other fields or sources
of positive ghost numbers are present in that case. This ensures that the
local contributions to the potential anomalies satisfy property ($ii^{\prime
}$) for $g=1$.

Condition ($iii$) is also suggested by the properties of renormalization.
If the regularization preserves the basic properties of the structure of $%
S_{\xi }$, then it is easy to show that the counterterms have the same
properties. In particular, we can arrange the regularization so that no
nontrivial one-particle irreducible diagrams can be constructed with
external legs of types $B$, $K_{\bar{C}}$ and $K_{B}$. It is even easier to
ensure that $X_{\xi }$ and $Y_{\xi }$ are independent of $\ul{K}$,
since the $\ul{K}$-dependent terms of the action are just made of
external fields and do not even need to be regularized.

To make some steps of the derivation clearer, we write $Y_{\xi }=Y_{\xi
}(\phi ,C,\bar{C},\ul{\phi },\ul{C},\tilde{K}_{\phi }(\xi
),K_{C},\xi )$, where $\tilde{K}_{\phi }(\xi )$ is given in (\ref{ktfi}). It
is not necessary to organize the variables of $X_{\xi }$ in a similar way,
so we just write $X_{\xi }=X_{\xi }(\Phi ,\ul{\Phi },K,\xi )$.

The assumption (\ref{i1}) is actually necessary for just one value of $\xi $%
, since then equation (\ref{i2}) implies (\ref{i1}) for every $\xi $.
Indeed, taking the derivative of $\llbracket  S_{\xi },X_{\xi }\rrbracket $ with respect to $%
\xi $ and using (\ref{i0}) and (\ref{i2}), we get%
\begin{equation}
\frac{\partial }{\partial \xi }\llbracket  S_{\xi },X_{\xi }\rrbracket -\llbracket  \llbracket  S_{\xi },X_{\xi
}\rrbracket ,\tilde{Q}\rrbracket =0.  \label{e1}
\end{equation}%
This equation can be easily integrated \cite{ABward,back,CBHcanon}. The
result is that the $\xi $ dependence of $\llbracket  S_{\xi },X_{\xi }\rrbracket $ is just due
to the canonical transformation generated by $\mathcal{C}(-\xi Q)$. Thus, if 
$\llbracket  S_{\xi },X_{\xi }\rrbracket $ vanishes for some $\xi $, it vanishes for all $\xi $%
.

\subsubsection*{Statement}

We want to prove that there exist local functionals $\mathcal{G}(\phi ,C)$
and $\chi _{\xi }(\Phi ,\ul{\Phi },K,\xi )$ such that%
\begin{equation}
X_{\xi }(\Phi ,\ul{\Phi },K,\xi )=\mathcal{G}(\phi +\ul{\phi }%
,C+\ul{C})+\llbracket  S_{\xi },\chi _{\xi }\rrbracket .  \label{thm}
\end{equation}%
Moreover, we show that $\chi _{\xi }$ is independent of $B$, $K_{\bar{C}}$
and $K_{B}$, while $\chi _{\xi }(\{0,0,\bar{C},B\},\ul{\Phi },K,1)=0$
and $\chi _{\xi }(\Phi ,\ul{\Phi },K,0)$ just depends on $\Phi +%
\ul{\Phi }$ and $K$. We also find the explicit expression of $\chi
_{\xi }$.

Note that $\mathcal{G}$ is independent of $\xi $. Taking the squared
antiparentheses $\llbracket  \cdots \rrbracket $ of both sides of (\ref{thm}) with $S_{\xi }$
and using (\ref{sm1}) and (\ref{i1}), we get $\llbracket  S_{\xi },\mathcal{G}\rrbracket =0$.
In the case $g=0$, the functional $\mathcal{G}$ is $C$ independent\ and
gauge invariant.

For convenience, we divide the proof into four steps.

\subsubsection*{Step 1 of proof: Interpolation between the background and
nonbackground sides}

Consider the functionals%
\begin{equation}
S_{R\xi }=S_{\xi }+wX_{\xi },\qquad \tilde{Q}_{R\xi }=\tilde{Q}+wY_{\xi },
\label{srxi}
\end{equation}%
where $w$ is a constant parameter such that $w^{2}=0$. If the ghost number $%
g $ of $X_{\xi }$ is odd, we can take $w=\varpi $, where $\varpi $ is
constant and anticommuting. If $g$ is even, we can take $w=\varpi \varpi
^{\prime }$, where $\varpi ^{\prime }$ is constant, anticommuting and
different from $\varpi $. We introduce\ the parameter $w$ to make the first
order of the $w$ Taylor expansion exact, which simplifies several arguments.
We have used the subscript R in (\ref{srxi}), because in several practical
applications $X_{\xi }$ and $Y_{\xi }$ are counterterms, while $S_{R\xi }$
and $\tilde{Q}_{R\xi }$ are renormalized functionals.

Combining $\llbracket  S_{\xi },S_{\xi }\rrbracket =0$ with (\ref{i1}) and (\ref{i0}) with (%
\ref{i2}), we get%
\begin{equation}
\llbracket  S_{R\xi },S_{R\xi }\rrbracket =0,\qquad \frac{\partial S_{R\xi }}{\partial \xi }%
=\llbracket  S_{R\xi },\tilde{Q}_{R\xi }\rrbracket ,  \label{speq}
\end{equation}%
which are the analogues of the equations in (\ref{gauind}). The second formula
implies that the canonical transformation generated by 
\begin{equation}
F_{\xi }(\Phi ,\ul{\Phi },K^{\prime },\ul{K}^{\prime },\xi )=%
\mathcal{C}(\xi Q)+w\int_{0}^{\xi }\mathrm{d}\xi ^{\prime }Y_{\xi }(\phi ,C,%
\bar{C},\ul{\phi },\ul{C},\tilde{K}_{\phi }^{\prime
},K_{C}^{\prime },\xi ^{\prime }),  \label{fr}
\end{equation}%
where%
\begin{equation}
\tilde{K}_{\phi }^{i\hspace{0.01in}\prime }=K_{\phi }^{i\hspace{0.01in}%
\prime }-\bar{C}^{I\hspace{0.01in}}G^{Ii}(0,-\overleftarrow{\partial }),
\label{kip}
\end{equation}%
interpolates between the nonbackground and background values of the action $%
S_{R\xi }$, which are $S_{\text{nb}}+wX_{0}\equiv S_{\text{nb}R}$ and $S_{%
\text{b}}+wX_{1}\equiv S_{\text{b}R}$, respectively. In compact notation, we
have%
\begin{equation}
S_{\text{nb}R}=S_{\text{nb}}+wX_{0}=F_{\xi }S_{R\xi }.  \label{inter2}
\end{equation}%
To check that (\ref{fr}) implies (\ref{speq}), it is sufficient to apply
formula (\ref{bu}) to (\ref{inter2}) and express $\partial F_{\xi }/\partial
\xi $ in terms of the fields and the sources without primes.

The composition formulas recalled in the appendix allow us to write the
relation%
\begin{equation}
F_{\xi }=F_{Y\xi }\mathcal{C}(\xi Q),  \label{compo}
\end{equation}%
where $F_{Y\xi }$ denotes the canonical transformation generated by 
\begin{equation}
F_{Y\xi }(\Phi ,\ul{\Phi },K^{\prime },\ul{K}^{\prime },\xi
)=\int \ \Phi ^{\alpha }K_{\alpha }^{\prime }+\int \ \ul{\Phi }%
^{\alpha }\ul{K}_{\alpha }^{\prime }+w\int_{0}^{\xi }\mathrm{d}\xi
^{\prime }Y_{\xi }(\phi ,C,\bar{C},\ul{\phi },\ul{C},\tilde{K}%
_{\phi }^{i\hspace{0.01in}\prime },K_{C}^{\prime },\xi ^{\prime }).
\label{fY}
\end{equation}%
Indeed, formula (\ref{fista}) reduces to $C=A+B$ in this case. All the
corrections vanish, since the nontrivial part of $\mathcal{C}(\xi Q)$ does
not depend on any sources $K^{\prime }$, $\ul{K}^{\prime }$ other
than $K_{B}^{\prime }$, while $Y_{\xi }$ is $B$ independent.

Now we compare the background and nonbackground \textquotedblleft
renormalized\textquotedblright\ actions $S_{\text{b}R}$ and $S_{\text{nb}R}$%
. Evaluating (\ref{inter2}) at $\xi =1$, we get the equality

\begin{equation}
\hat{S}_{\text{nb}R}(\Phi ^{\prime }+\ul{\Phi },K^{\prime })-\int
R^{\alpha }(\ul{\Phi })(\ul{K}_{\alpha }^{\prime }-K_{\alpha
}^{\prime })=\hat{S}_{\text{b}R}(\Phi ,\ul{\Phi },K)-\int \mathcal{R}%
^{\alpha }(\Phi ,\ul{C})K_{\alpha }-\int R^{\alpha }(\ul{\Phi }%
)\ul{K}_{\alpha },  \label{ei}
\end{equation}%
where the fields and the sources with primes are related to those without
primes by the canonical transformation $F_{\xi }$ with $\xi =1$, while 
\begin{eqnarray}
\hat{S}_{\text{nb}R}(\Phi +\ul{\Phi },K) &=&\hat{S}_{\text{nb}}(\Phi +%
\ul{\Phi },K)+wX_{0}(\Phi +\ul{\Phi },K),  \notag \\
\hat{S}_{\text{b}R}(\Phi ,\ul{\Phi },K) &=&\hat{S}_{\text{b}}(\Phi ,%
\ul{\phi },K)+wX_{1}(\Phi ,\ul{\Phi },K).  \label{fei}
\end{eqnarray}%
Using (\ref{passo}) and (\ref{fr}), we find, among other things, $\ul{%
\Phi }^{\alpha \hspace{0.01in}\prime }=\ul{\Phi }^{\alpha }$ and 
\begin{equation}
\phi ^{i\hspace{0.01in}\prime }=\phi ^{i}+\int_{0}^{1}\mathrm{d}\xi \frac{%
\delta (wY_{\xi })}{\delta K_{\phi }^{i\hspace{0.01in}\prime }},\quad C^{I%
\hspace{0.01in}\prime }=C^{I}+\int_{0}^{1}\mathrm{d}\xi \frac{\delta
(wY_{\xi })}{\delta K_{C}^{I\hspace{0.01in}\prime }},\quad \bar{C}^{I\hspace{%
0.01in}\prime }=\bar{C}^{I},\quad B^{I\hspace{0.01in}\prime }=B^{I}+\mathcal{%
R}_{\bar{C}}^{I}(\bar{C},\ul{C}).  \label{beside}
\end{equation}

\subsubsection*{Step 2 of proof: The background field trick}

The basic trick of the background field method is to switch the quantum
fields off and resume them later with the help of the background fields. Now
we explain how this trick is implemented when the background field method is
used together with the BV formalism, and we show how this helps us achieve our
goal.

First, we need to express both sides of equation (\ref{ei}) in terms of the
fields $\Phi $, $\ul{\Phi }$ and the sources $K^{\prime }$, $%
\ul{K}^{\prime }$. Once this is done, we set $\phi =C=\ul{K}%
^{\prime }=0$ and keep $\ul{\phi },\ul{C},\bar{C},B$ and $%
K^{\prime }$ as independent fields and sources. We denote the fields $\Phi
^{\alpha \hspace{0.01in}\prime }$ and the sources $K_{\alpha }$, $\ul{%
K}_{\alpha }$ obtained by applying these operations by $\Phi _{0}^{\alpha 
\hspace{0.01in}\prime }$ and $K_{\alpha 0}$, $\ul{K}_{\alpha 0}$,
respectively. We find%
\begin{eqnarray}
\hat{S}_{\text{nb}R}(\Phi _{0}^{\prime }+\ul{\Phi },K^{\prime }) &=&%
\hat{S}_{\text{b}R}(\{0,0,\bar{C},B\},\ul{\Phi },K_{0})  \notag \\
&&-\int \mathcal{R}_{\bar{C}}^{I}(\bar{C},\ul{C})K_{\bar{C}%
0}^{I}-\int \mathcal{R}_{B}^{I}(B,\ul{C})K_{B0}^{I}-\int R^{\alpha }(%
\ul{\Phi })(\ul{K}_{\alpha 0}+K_{\alpha }^{\prime }).
\label{min}
\end{eqnarray}%
Moreover, (\ref{fei}), (\ref{shat}) and assumption ($ii$) give 
\begin{equation}
\hat{S}_{\text{b}R}(\{0,0,\bar{C},B\},\ul{\Phi },K_{0})=S_{c}(%
\ul{\phi })+wX_{1}(0,\ul{\Phi },0)-\int B^{I}K_{\bar{C}%
0}^{I}+\int B^{I}\zeta _{IJ}(\ul{\phi },\partial )B^{J}.  \label{mina}
\end{equation}

Consider the canonical transformation $\{\ul{\phi },\ul{C},%
\bar{C},B\},\breve{K}\rightarrow \Phi ^{\prime \prime },K^{\prime }$ defined
by the generating functional 
\begin{eqnarray}
&&F(\{\ul{\phi },\ul{C},\bar{C},B\},K^{\prime }) =\int 
\ul{\phi }^{i}K_{\phi }^{i\hspace{0.004in}\prime }+\int \ \ul{C%
}^{I}K_{C}^{I\prime }+F_{\xi }(\{0,0,\bar{C},B\},\ul{\Phi },K^{\prime
},0,1)=\int \ul{\phi }^{i}K_{\phi }^{i\hspace{0.004in}\prime }+\int \ 
\ul{C}^{I}K_{C}^{I\prime }\hspace{-5pt}  \notag \\
&&+\int \bar{C}^{I}K_{\bar{C}}^{I\prime }+\int B^{I}K_{B}^{I\prime }-\int 
\bar{C}^{I}G^{Ii}(0,\partial )\ul{\phi }^{i}+\int \bar{C}^{I}(\zeta
_{IJ}(\ul{\phi },\partial )-\zeta _{IJ}(0,\partial ))B^{J}+\int 
\mathcal{R}_{\bar{C}}^{I}(\bar{C},\ul{C})K_{B}^{I\prime }\hspace{-5pt}
\notag \\
&&-\int \bar{C}^{I}\zeta _{IJ}(0,\partial )\mathcal{R}_{\bar{C}}^{J}(\bar{C},%
\ul{C})+w\int_{0}^{1}\mathrm{d}\xi \hspace{0.02in}Y_{\xi }(0,0,\bar{C}%
,\ul{\phi },\ul{C},\tilde{K}_{\phi }^{i\hspace{0.01in}\prime
},K_{C}^{\prime },\xi ).\qquad \qquad  \label{for}
\end{eqnarray}%
Using (\ref{beside}) and the other transformation rules not reported in that
formula, we find 
\begin{equation*}
\Phi ^{\alpha \hspace{0.01in}\hspace{0.01in}\prime \prime }=\ul{\Phi }%
^{\alpha }+\Phi _{0}^{\alpha \hspace{0.01in}\prime },\qquad \breve{K}_{\phi
}^{i}=K_{\phi }^{i\hspace{0.01in}\prime }+\ul{K}_{\phi 0}^{i},\qquad 
\breve{K}_{C}^{I}=K_{C}^{I\hspace{0.01in}\prime }+\ul{K}%
_{C0}^{I},\qquad \breve{K}_{\bar{C}}^{I}=K_{\bar{C}0}^{I},\qquad \breve{K}%
_{B}^{I}=K_{B0}^{I},
\end{equation*}%
which, with the help of (\ref{mina}), turn the identity (\ref{min})\ into 
\begin{eqnarray}
\hat{S}_{\text{nb}R}(\Phi ^{\prime \prime },K^{\prime }) &=&S_{c}(\ul{%
\phi })+\int B^{I}\zeta _{IJ}(\ul{\phi },\partial )B^{J}+wX_{1}(0,%
\ul{\Phi },0)-\int R_{\phi }^{i}(\ul{\Phi })\breve{K}_{\phi
}^{i}-\int R_{C}^{I}(\ul{\Phi })\breve{K}_{C}^{I}  \notag \\
&&-\int B^{I}\breve{K}_{\bar{C}}^{I}-\int \mathcal{R}_{\bar{C}}^{I}(\bar{C},%
\ul{C})\breve{K}_{\bar{C}}^{I}-\int \mathcal{R}_{B}^{I}(B,\ul{C%
})\breve{K}_{B}^{I}.  \label{fin0}
\end{eqnarray}

This is the key result we need. Now we elaborate it further and express it
in ways that make its contents more transparent.

\subsubsection*{Step 3 of proof: Simplification of the result}

First, we check that the master equation $(\hat{S}_{\text{nb}R},\hat{S}_{%
\text{nb}R})=0$ is satisfied. Assumption (\ref{i1}) at $\xi =1$ gives 
\begin{equation*}
0=\llbracket  S_{\text{b}},X_{1}\rrbracket =\int \left( \frac{\delta _{r}S_{\text{b}}}{\delta
\Phi ^{\alpha }}\frac{\delta _{l}X_{1}}{\delta K_{\alpha }}-\frac{\delta
_{r}S_{\text{b}}}{\delta K_{\alpha }}\frac{\delta _{l}X_{1}}{\delta \Phi
^{\alpha }}-\frac{\delta _{r}\bar{S}_{\text{b}}}{\delta \ul{K}%
_{\alpha }}\frac{\delta _{l}X_{1}}{\delta \ul{\Phi }^{\alpha }}%
\right) ,
\end{equation*}%
having used $\delta _{l}X_{1}/\delta \ul{K}_{\alpha }=0$ and $\delta
_{r}S_{\text{b}}/\delta \ul{K}_{\alpha }=\delta _{r}\bar{S}_{\text{b}%
}/\delta \ul{K}_{\alpha }$. Now, we set $\phi =C=0$ in this equation.
Recalling that $X_{1}$ is independent of $K$, $\bar{C}$ and $B$ at $\phi
=C=0 $, while $\delta _{r}S_{\text{b}}/\delta K_{\phi }^{i}$ and $\delta
_{r}S_{\text{b}}/\delta K_{C}^{I}$ vanish there, we get%
\begin{equation}
0=-\int \frac{\delta _{r}\bar{S}_{\text{b}}}{\delta \ul{K}_{\alpha }}%
\frac{\delta _{l}X_{1}(0,\ul{\Phi },0)}{\delta \ul{\Phi }%
^{\alpha }}=\llbracket  \bar{S}_{\text{b}},X_{1}(0,\ul{\Phi },0)\rrbracket ,
\label{gullo}
\end{equation}%
which states that $X_{1}(0,\ul{\Phi },0)$ is invariant under
background transformations. The conclusion obviously also applies to $S_{c}(%
\ul{\phi })$. The background invariance of the gauge fermion (\ref%
{psib}) implies that the second term on the right-hand side of (\ref{fin0})
is also invariant under background transformations. Thanks to these facts,
we can easily check, from the right-hand side of formula (\ref{fin0}), that $%
\hat{S}_{\text{nb}R}$ does satisfy the master equation $(\hat{S}_{\text{nb}%
R},\hat{S}_{\text{nb}R})=0$.

It is convenient to relabel the fields $\{\ul{\phi },\ul{C},%
\bar{C},B\}$ as $\Phi ^{\alpha }$, the sources $\breve{K}_{\alpha }$ as $%
K_{\alpha }$ and the fields $\Phi ^{\alpha \hspace{0.01in}\prime \prime }$
as $\Phi ^{\alpha \hspace{0.01in}\prime }$. Then formulas (\ref{for}) and (%
\ref{fin0}) tell us that the canonical transformation 
\begin{equation*}
F(\Phi ,K^{\prime })=\int \phi ^{i}K_{\phi }^{i\hspace{0.01in}\prime }+\int
\ C^{I}K_{C}^{I\hspace{0.01in}\prime }+F_{\xi }(\{0,0,\bar{C},B\},\{\phi
,C\},K^{\prime },0,1)
\end{equation*}%
is such that%
\begin{equation}
\tilde{F}_{\text{b}}^{-1}\hat{S}_{\text{nb}R}=F\hat{S}_{\text{b}R}^{\prime },
\label{side}
\end{equation}%
where 
\begin{eqnarray}
\hat{S}_{\text{b}R}^{\prime }(\Phi ,K) &=&S_{c}(\phi )+\int B^{I}\zeta
_{IJ}(\phi ,\partial )B^{J}+wX_{1}(0,\{\phi ,C\},0)-\int R^{\alpha }(\Phi
)K_{\alpha }  \notag \\
&&-\int \mathcal{R}_{\bar{C}}^{I}(\bar{C},C)K_{\bar{C}}^{I}-\int \mathcal{R}%
_{B}^{I}(B,C)K_{B}^{I},  \label{fin0ex}
\end{eqnarray}%
and $\tilde{F}_{\text{b}}^{-1}$ is the canonical transformation that undoes
the background shift. Precisely, $\tilde{F}_{\text{b}}^{-1}$ is the inverse
of the transformation generated by 
\begin{equation*}
\tilde{F}_{\text{b}}(\Phi ,\ul{\Phi },K^{\prime })=\int (\Phi
^{\alpha }-\ul{\Phi }^{\alpha })K_{\alpha }^{\prime },
\end{equation*}%
which is obtained from (\ref{fb}) by reducing the set of fields and sources
and downgrading the background fields $\ul{\Phi }$ to the role of
mere spectators.

Making the further canonical transformation $F_{\zeta }$, where 
\begin{equation*}
F_{\zeta }(\Phi ,K^{\prime })=\int \Phi ^{\alpha }K_{\alpha }^{\prime }+\int 
\bar{C}^{I}\zeta _{IJ}(\phi ,\partial )B^{J},\qquad
\end{equation*}%
we get $F_{\zeta }\hat{S}_{\text{b}R}^{\prime }=\hat{S}_{\text{b}R}^{\prime
\prime }$, where%
\begin{equation*}
\hat{S}_{\text{b}R}^{\prime \prime }(\Phi ,K)=S_{c}(\phi )+wX_{1}(0,\{\phi
,C\},0)-\int R^{\alpha }(\Phi )K_{\alpha }-\int \mathcal{R}_{\bar{C}}^{I}(%
\bar{C},C)K_{\bar{C}}^{I}-\int \mathcal{R}_{B}^{I}(B,C)K_{B}^{I}.
\end{equation*}%
In this derivation, it is helpful to recall that the last term of $F_{\zeta
} $ with $\phi \rightarrow \ul{\phi }$ is invariant under the
background transformations. Thus, $F_{\zeta }$ simply cancels the second
term on the right-hand side of (\ref{fin0ex}).

Finally, the canonical transformation generated by [check (\ref{fnm})]%
\begin{equation*}
\tilde{F}_{\text{nm}}(\Phi ,K^{\prime })=\int \Phi ^{\alpha }K_{\alpha
}^{\prime }-\int \mathcal{R}_{\bar{C}}^{I}(\bar{C},C)K_{B}^{\prime \hspace{%
0.01in}I},
\end{equation*}%
gives $\tilde{F}_{\text{nm}}^{-1}\hat{S}_{\text{b}R}^{\prime \prime }=\hat{S}%
_{\text{b}R}^{\prime \prime \prime }$, where%
\begin{equation*}
\hat{S}_{\text{b}R}^{\prime \prime \prime }(\Phi ,K)=S_{c}(\phi
)+wX_{1}(0,\{\phi ,C\},0)-\int R^{\alpha }(\Phi )K_{\alpha }=\mathcal{S}%
(\Phi ,K)+wX_{1}(0,\{\phi ,C\},0).
\end{equation*}%
In the last step, we have used (\ref{sfk}). Collecting the results found so
far, we get%
\begin{equation*}
\tilde{F}_{\text{b}}^{-1}\hat{S}_{\text{nb}R}=FF_{\zeta }^{-1}\tilde{F}_{%
\text{nm}}\hat{S}_{\text{b}R}^{\prime \prime \prime }.
\end{equation*}%
More explicitly, the canonical transformation $\Phi ,K\rightarrow \Phi
^{\prime },K^{\prime }$ generated by $FF_{\zeta }^{-1}\tilde{F}_{\text{nm}}$
is such that%
\begin{equation}
\hat{S}_{\text{nb}R}(\Phi ^{\prime },K^{\prime })=\mathcal{S}(\Phi
,K)+wX_{1}(0,\{\phi ,C\},0).  \label{keynb}
\end{equation}%
Note that formula (\ref{gullo}) can be recast in the form%
\begin{equation}
(\mathcal{S},X_{1}(0,\{\phi ,C\},0))=0.  \label{gul}
\end{equation}%
Together with $(\mathcal{S},\mathcal{S})=0$, this formula and (\ref{keynb})
give back $(\hat{S}_{\text{nb}R},\hat{S}_{\text{nb}R})=0$.

Let $\mathcal{G}_{i}(\phi ,C)$ denote a basis of local functionals of ghost
number $g$ that satisfy $(\mathcal{S},\mathcal{G}_{i})=0$ and are
constructed with the physical fields $\phi $ and the ghosts $C$. For $g=0$
they are just the usual gauge invariant local functionals. We have the
expansion 
\begin{equation}
X_{1}(0,\{\phi ,C\},0)=\sum_{i}\tau _{i}\mathcal{G}_{i}(\phi ,C),
\label{comple}
\end{equation}%
where $\tau _{i}$ are constants, and formula (\ref{keynb}) gives%
\begin{equation}
\hat{S}_{\text{nb}}(\Phi ^{\prime },K^{\prime })+wX_{0}(\Phi ^{\prime
},K^{\prime })=\mathcal{S}(\Phi ,K)+w\sum_{i}\tau _{i}\mathcal{G}_{i}(\phi
,C).  \label{eqq}
\end{equation}

This formula relates the background field functionals with the nonbackground
functionals. With the help of a few other manipulations, we can work out the
interpolation between the two sides and conclude the proof of the theorem.

\subsubsection*{Step 4 of proof: Interpolation of the result}

Using (\ref{passo}) and (\ref{fr}), it is easy to check the composition rule 
$F=\bar{F}_{Y}\bar{F}$, where $F$ is the transformation encoded in (\ref{for}%
), while $\bar{F}$ and $\bar{F}_{Y}$ are the transformations generated by 
\begin{eqnarray}
\bar{F}(\Phi ,K^{\prime }) &=&\int \Phi ^{\alpha }K_{\alpha }^{\prime }-\int 
\bar{C}^{I}G^{Ii}(0,\partial )\phi ^{i}+\int \bar{C}^{I}(\zeta _{IJ}(\phi
,\partial )-\zeta _{IJ}(0,\partial ))B^{J}  \notag \\
&&+\int \mathcal{R}_{\bar{C}}^{I}(\bar{C},C)K_{B}^{I\hspace{0.01in}\prime
}-\int \bar{C}^{I}\zeta _{IJ}(0,\partial )\mathcal{R}_{\bar{C}}^{J}(\bar{C}%
,C),  \label{fbar} \\
\bar{F}_{Y}(\Phi ,K^{\prime }) &=&\int \Phi ^{\alpha }K_{\alpha }^{\prime
}+w\int_{0}^{1}\mathrm{d}\xi \hspace{0.01in}Y_{\xi }(0,0,\bar{C},\phi ,C,%
\tilde{K}_{\phi }^{i\hspace{0.01in}\prime },K_{C}^{\prime },\xi ),
\label{fbary}
\end{eqnarray}%
respectively. Using this property and collecting the canonical
transformations made so far, we can write the identity (\ref{eqq}) in the
compact form%
\begin{equation}
\tilde{F}_{\text{nm}}^{-1}F_{\zeta }\bar{F}^{-1}\bar{F}_{Y}^{-1}\tilde{F}_{%
\text{b}}^{-1}(\hat{S}_{\text{nb}}+wX_{0})=\mathcal{S}+w\sum_{i}\tau _{i}%
\mathcal{G}_{i}.  \label{thus}
\end{equation}%
Applying the composition rules recalled in the appendix, in particular
formula (\ref{thispa}), it is easy to check that 
\begin{equation*}
\bar{F}F_{\zeta }^{-1}\tilde{F}_{\text{nm}}=\tilde{F}_{\text{gf}}^{\prime },
\end{equation*}%
where%
\begin{equation*}
\tilde{F}_{\text{gf}}^{\prime \hspace{0.01in}}(\Phi ,K^{\prime })=\int \
\Phi ^{\alpha }K_{\alpha }^{\prime }-\Psi ^{\prime }(\Phi )
\end{equation*}%
is the reduced version of (\ref{fgfp}). Thus, formula (\ref{thus})
simplifies to%
\begin{equation}
\tilde{F}_{\text{gf}}^{\prime \hspace{0.01in}-1}\bar{F}_{Y}^{-1}\tilde{F}_{%
\text{b}}^{-1}(\hat{S}_{\text{nb}}+wX_{0})=\mathcal{S}+w\sum_{i}\tau _{i}%
\mathcal{G}_{i}.  \label{this}
\end{equation}%
The terms independent of $w$ give 
\begin{equation}
\tilde{F}_{\text{gf}}^{\prime \hspace{0.01in}-1}\tilde{F}_{\text{b}}^{-1}%
\hat{S}_{\text{nb}}=\mathcal{S},  \label{esse}
\end{equation}%
which can be easily verified. Indeed, by formula (\ref{shatp}), if we undo
the background shift and the gauge fixing on the action $\hat{S}_{\text{nb}}$%
, we obtain the starting action $\mathcal{S}$.

The terms of (\ref{this}) that are proportional to $w$ can be better read
after acting on both sides with $\bar{F}_{Y}\tilde{F}_{\text{gf}}^{\prime }$%
, which gives%
\begin{equation}
w\tilde{F}_{\text{b}}^{-1}X_{0}=(\bar{F}_{Y}-1)\tilde{F}_{\text{b}}^{-1}\hat{%
S}_{\text{nb}}+w\sum_{i}\tau _{i}\mathcal{G}_{i}.  \label{gro}
\end{equation}%
Define the functional%
\begin{equation}
\Upsilon (\Phi ,K)=(-1)^{g+1}\int_{0}^{1}\mathrm{d}\xi ^{\prime }\hspace{%
0.01in}Y_{\xi }(0,0,\bar{C},\phi ,C,\bar{K}_{\phi },K_{C},\xi ^{\prime }),
\label{y}
\end{equation}%
where $\bar{K}_{\phi }$ is defined in formula (\ref{kbar}). Then from (\ref{gro}) we can write%
\begin{equation}
\tilde{F}_{\text{b}}^{-1}X_{0}=\sum_{i}\tau _{i}\mathcal{G}_{i}+(\tilde{F}_{%
\text{b}}^{-1}\hat{S}_{\text{nb}},\Upsilon ).  \label{x0}
\end{equation}

At this point, we start looking at the functionals as functionals of $\Phi ,%
\ul{\Phi },K,\ul{K}$ again. We obviously have $\tilde{F}_{%
\text{b}}^{-1}\hat{S}_{\text{nb}}=F_{\text{b}}^{-1}\hat{S}_{\text{nb}}$ and $%
\tilde{F}_{\text{b}}^{-1}X_{0}=F_{\text{b}}^{-1}X_{0}$, since $\hat{S}_{%
\text{nb}}$ and $X_{0}$ are independent of $\ul{K}$. Moreover, $(%
\tilde{F}_{\text{b}}^{-1}\hat{S}_{\text{nb}},\Upsilon )=\llbracket  F_{\text{b}}^{-1}%
\hat{S}_{\text{nb}},\Upsilon \rrbracket =\llbracket  F_{\text{b}}^{-1}S_{\text{nb}},\Upsilon
\rrbracket =F_{\text{b}}^{-1}\llbracket  S_{\text{nb}},F_{\text{b}}\Upsilon \rrbracket $, where we have
used $\llbracket  F_{\text{b}}^{-1}\bar{S}_{\text{nb}},\Upsilon \rrbracket =0$, which is due
to the fact that $F_{\text{b}}^{-1}\bar{S}_{\text{nb}}$ just depends on $%
\ul{\Phi },\ul{K}$, while $\Upsilon $ just depends on $\Phi ,K$%
. Applying $F_{\text{b}}$ to both sides of (\ref{x0}), we obtain 
\begin{equation}
X_{0}(\Phi +\ul{\Phi },K)=\sum_{i}\tau _{i}\mathcal{G}_{i}(\phi +%
\ul{\phi },C+\ul{C})+\llbracket  S_{\text{nb}},F_{\text{b}}\Upsilon \rrbracket .
\label{x1}
\end{equation}%
Using this result together with formula (\ref{inter2}), we find%
\begin{equation*}
F_{\xi }S_{R\xi }=S_{\text{nb}}+w\sum_{i}\tau _{i}\mathcal{G}_{i}(\phi +%
\ul{\phi },C+\ul{C})+w\llbracket  F_{\xi }S_{R\xi },F_{\text{b}%
}\Upsilon \rrbracket .
\end{equation*}%
Now we apply $F_{\xi }^{-1}$ to both sides and use $S_{R\xi }=S_{\xi
}+wX_{\xi }$ from the first formula of (\ref{srxi}) and $F_{\xi }^{-1}=%
\mathcal{C}(-\xi Q)F_{Y\xi }^{-1}$ from (\ref{compo}). Noting that $\mathcal{%
C}(-\xi Q)$ leaves $\mathcal{G}_{i}(\phi +\ul{\phi },C+\ul{C})$
invariant, because it does not affect the fields besides $B^{I}$, we find%
\begin{equation}
S_{\xi }+wX_{\xi }=\mathcal{C}(-\xi Q)F_{Y\xi }^{-1}S_{\text{nb}%
}+w\sum_{i}\tau _{i}\mathcal{G}_{i}(\phi +\ul{\phi },C+\ul{C}%
)+w\llbracket  S_{\xi },\mathcal{C}(-\xi Q)F_{\text{b}}\Upsilon \rrbracket .  \label{x2}
\end{equation}%
Using (\ref{fY}) and (\ref{inter}), we observe that 
\begin{equation*}
F_{Y\xi }^{-1}S_{\text{nb}}=S_{\text{nb}}-w\llbracket  S_{\text{nb}},\Upsilon _{\xi
}\rrbracket ,\qquad \mathcal{C}(-\xi Q)F_{Y\xi }^{-1}S_{\text{nb}}=S_{\xi }-w\llbracket 
S_{\xi },\mathcal{C}(-\xi Q)\Upsilon _{\xi }\rrbracket ,
\end{equation*}%
where 
\begin{equation}
\Upsilon _{\xi }(\Phi ,\ul{\Phi },K,\xi )=(-1)^{g+1}\int_{0}^{\xi }%
\mathrm{d}\xi ^{\prime }Y_{\xi }(\phi ,C,\bar{C},\ul{\phi },%
\ul{C},\bar{K}_{\phi },K_{C},\xi ^{\prime }).  \label{yx}
\end{equation}%
Finally, formula (\ref{x2}) gives%
\begin{equation}
X_{\xi }(\Phi ,\ul{\Phi },K,\xi )=\sum_{i}\tau _{i}\mathcal{G}%
_{i}(\phi +\ul{\phi },C+\ul{C})+\llbracket  S_{\xi },\chi _{\xi }\rrbracket ,
\label{thmf}
\end{equation}%
where%
\begin{equation}
\chi _{\xi }(\Phi ,\ul{\Phi },K,\xi )=\mathcal{C}(-\xi Q)\left( F_{%
\text{b}}\Upsilon -\Upsilon _{\xi }\right) .  \label{chix}
\end{equation}

Formula (\ref{thmf}) coincides with (\ref{thm}) once we set $\mathcal{G}%
(\phi ,C)=\sum_{i}\tau _{i}\mathcal{G}_{i}(\phi ,C)$. Using (\ref{passo}), (%
\ref{y}), (\ref{yx}) and (\ref{ktfi}), we find%
\begin{eqnarray*}
\chi _{\xi }(\Phi ,\ul{\Phi },K,\xi ) &=&(-1)^{g+1}\int_{0}^{1}%
\mathrm{d}\xi ^{\prime }\hspace{0.01in}Y_{\xi }(0,0,\bar{C},\phi +\ul{%
\phi },C+\ul{C},\tilde{K}_{\phi }(\xi ),K_{C},\xi ^{\prime }) \\
&&-(-1)^{g+1}\int_{0}^{\xi }\mathrm{d}\xi ^{\prime }Y_{\xi }(\phi ,C,\bar{C},%
\ul{\phi },\ul{C},\tilde{K}_{\phi }(\xi ),K_{C},\xi ^{\prime
}).
\end{eqnarray*}%
This functional is local and independent of $B$, $K_{\bar{C}}$, $K_{B}$ and $%
\ul{K}$. Moreover, $\chi _{\xi }(\Phi ,\ul{\Phi },K,0)$ just
depends on $\Phi +\ul{\Phi }$ and $K$. Finally, $\chi _{\xi }(\{0,0,%
\bar{C},B\},\ul{\Phi },K,1)=0$. This concludes the proof of the
theorem.

Observe that $\chi _{\xi }$ satisfies the equation%
\begin{equation*}
\frac{\partial \chi _{\xi }}{\partial \xi }-\llbracket  \chi _{\xi },\tilde{Q}%
\rrbracket =(-1)^{g}Y_{\xi },
\end{equation*}%
which can be proved by using (\ref{bu}) on (\ref{chix}) and noting that $%
\mathcal{C}(-\xi Q)\tilde{Q}=\tilde{Q}$.

Formulas (\ref{thmf}) and (\ref{comple}) tell us that, in the end, the
nontrivial cohomological content of the functional $X_{\xi }$ is just $%
X_{1}(0,\ul{\Phi },0)$, which can be worked out with the background
field method by evaluating Feynman diagrams that just have background fields 
$\ul{\phi }$ and background ghosts $\ul{C}$ on their external
legs.

\section{Conclusions}

\setcounter{equation}{0}

\label{s4}

We have studied the cohomology of the local functionals of arbitrary ghost
numbers generated by renormalization in quantum field theories whose gauge
symmetries are general covariance, local Lorentz symmetry, non-Abelian
Yang-Mills symmetries and Abelian gauge symmetries. The case of ghost number 0
is important to characterize the divergent parts of Feynman diagrams.
The case of ghost number 1 is important for the local contributions to the
potential anomalies.

Using the Batalin-Vilkovisky formalism and the background field method, we
have proved that a closed local functional can always be written as a local
functional that just depends on the physical fields $\phi $ and the
Faddeev-Popov ghosts $C$, plus an exact functional, which is equal to the BV
antiparentheses between the action and another local functional. Every term
that depends nontrivially on the antighosts $\bar{C}$, the Lagrange
multipliers $B$ for the gauge fixing and/or the sources $K$ coupled to the
symmetry transformations of the fields is cohomologically trivial.

The basic idea of the proof is to interpolate between the background field
approach and the usual, nonbackground approach by means of canonical
transformations. This allows us to take advantage of the virtues of both
approaches, and highlight, among other things, that the counterterms and the
local contributions to the potential anomalies are not just cohomologically
closed, but satisfy more restrictive conditions. We have managed to
translate those conditions into simple mathematical assumptions and obtain a
general theorem. The result supersedes numerous involved arguments that
exist in the literature and offers a better understanding of the matter. It
can be used to upgrade the recent proof \cite{ABnonreno} of the
Adler-Bardeen theorem in nonrenormalizable theories.

\renewcommand{\thesection}{A}

\section*{Appendix. Useful formulas}

\label{appA}

\setcounter{equation}{0}\renewcommand{\theequation}{\thesection.%
\arabic{equation}}

In this appendix we collect a few reference formulas that are used in the
paper.

The functional $-\int R^{\alpha }(\Phi )K_{\alpha }$ that appears in (\ref%
{sfk}) reads 
\begin{eqnarray*}
&&\int (C^{\rho }\partial _{\rho }A_{\mu }^{a}+A_{\rho }^{a}\partial _{\mu
}C^{\rho }-\partial _{\mu }C^{a}-gf^{abc}A_{\mu }^{b}C^{c})K_{A}^{\mu
a}+\int \left( C^{\rho }\partial _{\rho }C^{a}+\frac{g}{2}%
f^{abc}C^{b}C^{c}\right) K_{C}^{a} \\
&&+\int (C^{\rho }\partial _{\rho }e_{\mu }^{\hat{a}}+e_{\rho }^{\hat{a}%
}\partial _{\mu }C^{\rho }+C^{\hat{a}\hat{b}}e_{\mu \hat{b}})K_{\hat{a}%
}^{\mu }+\int C^{\rho }(\partial _{\rho }C^{\mu })K_{\mu }^{C}+\int (C^{\hat{%
a}\hat{c}}\eta _{\hat{c}\hat{d}}C^{\hat{d}\hat{b}}+C^{\rho }\partial _{\rho
}C^{\hat{a}\hat{b}})K_{\hat{a}\hat{b}}^{C} \\
&&\!\!\!\!\!\!{+\int \left( C^{\rho }\partial _{\rho }\bar{\psi}_{L}-\frac{i%
}{4}\bar{\psi}_{L}\sigma ^{\hat{a}\hat{b}}C_{\hat{a}\hat{b}}+g\bar{\psi}%
_{L}T^{a}C^{a}\right) K_{\psi }+\int K_{\bar{\psi}}\left( C^{\rho }\partial
_{\rho }\psi _{L}-\frac{i}{4}\sigma ^{\hat{a}\hat{b}}C_{\hat{a}\hat{b}}\psi
_{L}+gT^{a}C^{a}\psi _{L}\right) } \\
&&+\int \left( C^{\rho }(\partial _{\rho }\varphi )+g\mathcal{T}%
^{a}C^{a}\varphi \right) K_{\varphi }-\int B^{a}K_{\bar{C}}^{a}-\int B_{\mu
}K_{\bar{C}}^{\mu }-\int B_{\hat{a}\hat{b}}K_{\bar{C}}^{\hat{a}\hat{b}}.
\end{eqnarray*}%
Here $\hat{a},\hat{b},\ldots $ are local Lorentz indices, while $C^{\mu }$-$%
\bar{C}_{\mu }$-$B_{\mu }$, $C^{\hat{a}\hat{b}}$-$\bar{C}_{\hat{a}\hat{b}}$-$%
B^{\hat{a}\hat{b}}$ and $C^{a}$-$\bar{C}^{a}$-$B^{a}$ denote the ghosts, the
antighosts and the Lagrange multipliers of diffeomorphisms, local Lorentz
symmetry and Yang-Mills symmetry, respectively. Moreover, $\varphi $ are
scalar fields and $\psi _{L}$ are left-handed fermions, while $\mathcal{T}%
^{a}$ and $T^{a}$ are the anti-Hermitian matrices\ associated with their
representations.

A helpful identity tells us that if $Y(\Phi ,K)$ is a functional that
behaves as a scalar [i.e. such that $Y^{\prime }(\Phi ^{\prime },K^{\prime
})=Y(\Phi ,K)$] under the canonical transformation $\Phi ,K\rightarrow \Phi
^{\prime },K^{\prime }$, generated by the functional $F(\Phi ,K^{\prime })$,
then we have \cite{removal,back} 
\begin{equation}
\frac{\partial Y^{\prime }}{\partial \zeta }=\frac{\partial Y}{\partial
\zeta }-(Y,\tilde{F}_{\zeta }),  \label{bu}
\end{equation}%
where $\tilde{F}_{\zeta }(\Phi ,K)=F_{\zeta }(\Phi ,K^{\prime }(\Phi ,K))$
and $F_{\zeta }(\Phi ,K^{\prime })=\partial F/\partial \zeta $. In this
formula, the $\zeta $ derivative of each functional is evaluated while the
natural arguments of the functional are kept constant. Precisely, $\Phi
^{\prime }$ and $K^{\prime }$ are constant in the $\zeta $ derivative of $%
Y^{\prime }$, while $\Phi $ and $K$ are constant in the $\zeta $ derivative
of $Y$, and $\Phi $ and $K^{\prime }$ are constant in the $\zeta $
derivative of $F$.

In ref. \cite{CBHcanon}, various formulas for the manipulation of canonical
transformations at the level of their generating functions have been given.
In particular, if $F_{A}(q,P)=q^{i}P^{i}+A(q,P)$ and $%
F_{B}(q,P)=q^{i}P^{i}+B(q,P)$ are the generating functions of two canonical
transformations $q,p\rightarrow Q,P$, the generating function of the
composed transformation $F_{C}=F_{B}F_{A}$ is written as $%
F_{C}(q,P)=q^{i}P^{i}+C(q,P)$, where the function $C(q,P)$ is expressed as a
sum of monomials built with $A$, $B$ and their derivatives%
\begin{equation*}
A_{i_{1}\cdots i_{n}}=\frac{\partial ^{n}A(q,P)}{\partial P_{i_{1}}\cdots
\partial P_{i_{n}}},\qquad B^{i_{1}\cdots i_{n}}=\frac{\partial ^{n}B(q,P)}{%
\partial q_{i_{1}}\cdots \partial q_{i_{n}}}.
\end{equation*}%
The first contributions to $C(q,P)$ are%
\begin{equation}
C=A+B+A_{i}B^{i}+\frac{1}{2}A_{i}B^{ij}A_{j}+\frac{1}{2}B^{i}A_{ij}B^{j}+%
\cdots  \label{fista}
\end{equation}

These results extend straightforwardly to the BV formalism. When we compose
the canonical transformations $\Phi ,K\rightarrow \Phi ^{\prime },K^{\prime
} $ generated by $F_{A}(\Phi ,K^{\prime })=\int \Phi ^{\alpha }K_{\alpha
}^{\prime }+A(\Phi ,K^{\prime })$ and $F_{B}(\Phi ,K^{\prime })=\int \Phi
^{\alpha }K_{\alpha }^{\prime }+B(\Phi ,K^{\prime })$, we write the result
as $F_{C}=F_{B}F_{A}$, where $F_{C}(\Phi ,K^{\prime })=\int \Phi ^{\alpha
}K_{\alpha }^{\prime }+C(\Phi ,K^{\prime })$. In the applications of this
paper, due to the simple structures of the functionals $A$ and $B$, formula (%
\ref{fista}) effectively reduces to 
\begin{equation}
C=A+B+\int \frac{\delta A}{\delta K_{\alpha }^{\prime }}\frac{\delta B}{%
\delta \Phi ^{\alpha }}.  \label{thispa}
\end{equation}

Several operations on canonical transformations can be handled more
practically by means of the componential map $\mathcal{C}$ \cite{CBHcanon},
which expresses the generating function $F(q,P)$ in terms of a another
function $X(q,P)$ as $\mathcal{C}(X(q,P))$, with the expansion%
\begin{equation}
\mathcal{C}(X)=I+X+\frac{1}{2}X_{i}X^{i}+\frac{1}{3!}\left(
X_{ij}X^{i}X^{j}+X^{j}X_{j}^{i}X_{i}+X^{ij}X_{i}X_{j}\right) +\cdots ,
\label{expac}
\end{equation}%
where 
\begin{equation*}
X_{j_{i}\cdots j_{m}}^{i_{1}\cdots i_{n}}\equiv \frac{\partial ^{n+m}X(q,P)}{%
\partial q^{i_{1}}\cdots \partial q^{i_{n}}\partial P^{j_{1}}\cdots \partial
P^{j_{m}}}.
\end{equation*}%
The advantage of the componential map $\mathcal{C}(X)$ is that it satisfies
the Baker-Campbell-Hausdorff (BCH) formula, like the exponential map $%
\mathrm{e}^{\mathrm{ad}(X)}$, where $\mathrm{ad}(X)Y=\{X,Y\}$ is the adjoint
map and $\{X,Y\}$ are the Poisson brackets of $X$ and $Y$. Precisely, if we
define $X\triangle Y$ from the BCH formula 
\begin{equation}
\mathrm{e}^{\mathrm{ad}(X)}\mathrm{e}^{\mathrm{ad}(Y)}=\mathrm{e}^{\mathrm{ad%
}(X+Y+X\triangle Y)},  \label{bosoes}
\end{equation}%
then the componential map satisfies%
\begin{equation*}
\mathcal{C}(X)\mathcal{C}(Y)=\mathcal{C}(X+Y+X\triangle Y),
\end{equation*}%
where the product on the left-hand side is the composition of the canonical
transformations. In particular, the inverse of $\mathcal{C}(X)$ is just $%
\mathcal{C}(-X)$.

Again, the generalization to the BV formalism is straightforward and, due to
the simple structures of the functionals $X(\Phi ,K^{\prime })$, in most
applications of this paper the expansion (\ref{expac}) of the componential
map effectively reduces to%
\begin{equation}
\mathcal{C}(X(\Phi ,K^{\prime }))=\int \Phi ^{\alpha }K_{\alpha }^{\prime
}+X(\Phi ,K^{\prime })+\frac{1}{2}\int \frac{\delta X}{\delta K_{\alpha
}^{\prime }}\frac{\delta X}{\delta \Phi ^{\alpha }}.  \label{boso}
\end{equation}

\end{document}